\documentclass[aps,preprint,showpacs,preprintnumbers,amsmath,amssymb]{revtex4}
\usepackage{amsmath,mathrsfs,amsbsy,color,graphicx,bm,amsthm,amsfonts}
\usepackage{units}
\usepackage{bbm}
\usepackage{times}
\usepackage{dcolumn}
\usepackage{mathrsfs}
\usepackage{amsmath,amssymb,epsfig}
\usepackage{amsmath}
\newcommand{\udots}{\mathinner{\mskip1mu\raise1pt\vbox{\kern7pt\hbox{.}}
\mskip2mu\raise4pt\hbox{.}\mskip2mu\raise7pt\hbox{.}\mskip1mu}}
\begin{document}
\title{ Does anti-Unruh effect assist quantum entanglement and coherence?}
\author{Shu-Min Wu$^1$\footnote{Email: smwu@lnnu.edu.cn}, Xiao-Wei Teng$^1$, Jin-Xuan Li$^1$, Hao-Sheng Zeng$^2$\footnote{Email: hszeng@hunnu.edu.cn}, Tonghua Liu$^3$\footnote{Email:liutongh@yangtzeu.edu.cn (corresponding author)} }
\affiliation{$^1$ Department of Physics, Liaoning Normal University, Dalian 116029, China\\
$^2$ Department of Physics, Hunan Normal University, Changsha 410081, China\\
$^3$ School of Physics and Optoelectronic Engineering, Yangtze University, Jingzhou 434023
}


\begin{abstract}
In this paper, we use the concepts of quantum entanglement and coherence to analyze the Unruh and anti-Unruh effects based on the model of Unruh-DeWitt detector. For the first time, we find that (i) the Unruh effect reduces quantum entanglement but enhances quantum coherence; (ii) the anti-Unruh effect enhances quantum entanglement but reduces quantum coherence.
This surprising result refutes the notion that the Unruh effect can only destroy quantum entanglement and coherence simultaneously, and that the anti-Unruh can only protect quantum resources. Consequently,  it opens up a new source for discovering experimental evidence supporting the existence of the Unruh and anti-Unruh effects.
\end{abstract}

\vspace*{0.5cm}
\maketitle
\section{Introduction}
Quantum entanglement plays an important role in quantum information science. It is a necessary ingredient for various computational tasks, such as quantum remote control, quantum teleportation and quantum communication \cite{L1,L2,L3}. A lot of  progresses have been made in understanding  the behavior of quantum entanglement in various aspects, such as the sudden death and sudden birth, the degeneration or enhancement of quantum entanglement \cite{L4,L5,L6,L7,L8,L9,L10}. On the other hand, as a broader concept, quantum coherence is also a physical resource in quantum technologies, optical experiments and biological systems \cite{L11,L12,L13,L14,L15}.
Many works have been done about how the environment influences quantum coherence and how to protect it \cite{L16,L17,L18,L19}. Quantum entanglement and coherence are closely related to each other. Generally speaking, quantum coherence is a necessary condition for quantum entanglement. Despite considerable efforts dedicated to investigating the relationship between quantum entanglement and coherence \cite{L20,AA2,L21,L22,AA,AA1}, several challenges still remain unresolved.

The Unruh effect, first proposed by Unruh in 1976 \cite{L23,L24}, stands as a crucial prediction of quantum field theory.
An inertial observer which undergoes uniform acceleration in the Minkowski vacuum will detect a thermal bath of particles of a free quantum field with a temperature proportional to the acceleration.
On the other hand, Hawking discovered that black holes can emit thermal radiation, a phenomenon subsequently named Hawking radiation \cite{L25}. According to the equivalence principle, the investigation of the Unruh effect holds great significance for studying Hawking radiation and its associated issues, including thermodynamics and the problem of information loss \cite{L26,L27,L28}.
Generally, both the Unruh effect and Hawking radiation have been observed to reduce quantum entanglement and coherence \cite{L29,L30,L31,L32,L33,L34,L35,L36,SMW1,SMW2,SMW3,SMW4,SMW5,SMW6,SMW7}.

Recent research has suggested the existence of the anti-Unruh effect, wherein, under specific conditions, the acceleration effect can also cool down a detector \cite{L37,L38,L39,qsc1,qsc2}. The concept of anti-Hawking radiation was also proposed \cite{L40}. For the global free models in the full space, the anti-Unruh effect cannot be detected physically. In order to observe the anti-Unruh effect, the semiclassical Unruh-DeWitt (UDW) detector model is usually employed, which consists of a two-level atom interacting locally with the vacuum field, and avoids the physically unfeasible detection of global models. The UDW detector model is commonly established in experiments, often within a finite length of optical cavity.
The results have demonstrated that the anti-Unruh effect enhances quantum entanglement for an initially entangled bipartite state \cite{L38,L39}.  However, the influence of the anti-Unruh effect on quantum coherence remains unclear. Additionally, it raises the question of whether the Unruh and anti-Unruh effects exert similar influences on both quantum entanglement and coherence within the UDW detector model. In previous studies \cite{L29,L30,L31,L32,L33,L34,L35,L36,SMW1,SMW2,SMW3,SMW4,SMW5,SMW6,SMW7}, both the free field model and the UDW model did not take boundary conditions into account. However, in our paper, we consider the boundary conditions in the UDW model. Through the investigation of these models, we may derive some intriguing conclusions, particularly highlighting how quantum correlations and coherence may exhibit distinctive properties under the influence of acceleration effects.

In this paper, we study the influence of acceleration effect on quantum entanglement and coherence. Assume that a spin qubit and a two-level atom is initially in an entangled pure state. The atom is then injected into a vacuum cavity with finite length and moves at a uniform acceleration along the length direction of the cavity. The atom plays the role of a detector which can detect the thermal radiation due to the acceleration effect. We want to know how the acceleration effect affects quantum entanglement and coherence.
The underlying motivation is to uncover novel aspects of the Unruh effect and anti-Unruh effect, contributing to a more comprehensive understanding of these phenomena.

The paper is organized as follows. In Sec. II, we briefly introduce the UDW detector model. In Sec. III, we study the influence of acceleration effect on quantum entanglement and coherence based on the UDW detector model.  The last section is devoted to the brief conclusion.

\section{Unruh-DeWitt model, \label{GSCDGE1}}
Let us first briefly recall the UDW model and the concept of anti-Unruh effect. The UDW model consists of a two-level atom (detector) interacting locally with a massless field $\phi(x(\tau))$ along the trajectory $x(\tau)$ with $\tau$ the detector's proper time \cite{L37}.
The detector has ground state $|g\rangle$ and excited state $|e\rangle$, which are separated by an energy gap $\Omega$. Suppose that the detector moves in a flat static cylinder with a spatial circumference ($L>0$).
This cylinder topology imposes periodic boundary conditions which is relevant to laboratory systems, such as the closed optical cavities, superconducting circuits coupled to periodic microwave guides and optical-fibre loops \cite{L41,L42,L43}.

In the interaction picture, the UDW Hamiltonian that describes the interaction between the detector and the field $\phi(x(\tau))$ is
\begin{eqnarray}\label{Q1}
H_I=\lambda\chi(\tau/\sigma)(e^{i\Omega\tau}\sigma^++e^{-i\Omega\tau}\sigma^-)\phi(x(\tau)),
\end{eqnarray}
where $\lambda$ is the coupling strength that is assume to be weak, $\sigma^\pm$ denote the ladder operators of detector, and $\chi(\tau/\sigma)$ is the switching function which controls the duration of interaction via the parameter $\sigma$.
The most natural choice for the switching function is the Gaussian function
\begin{eqnarray}\label{Q2}
\chi(\tau/\sigma)=e^{-\tau^2/2\sigma^2}.
\end{eqnarray}

For weak coupling, the unitary evolution of the total quantum system is given by \cite{L37}
\begin{eqnarray}\label{Q3}
U&=&\mathbb{I}+U^{(1)}+\mathcal{O}(\lambda^2)=\mathbb{I}-i\int d\tau H_I(\tau)+\mathcal{O}(\lambda^2)\\ \nonumber
&=&-i\lambda\sum_{m}(I_{+,m}{a}_{m}^{\dagger}\sigma^{+}+I_{-,m}{a}_{m}^{\dagger}\sigma^{-}+
\text{H.c.})+\mathcal{O}(\lambda^2),
\end{eqnarray}
where  $m$ denotes the mode of the  scalar field with annihilation and creation operators
$a_m|0\rangle=0$ and $a_m^\dag|0\rangle=|1_m\rangle$. The sum over $m$ takes discrete values due to the periodic boundary condition $k=2\pi m/L$, and $I_{\pm,m}$ can be written as
\begin{eqnarray}\label{Q6}
I_{\pm,m}=\int_{-\infty}^{\infty}\chi(\tau/\sigma) e^{\pm i\Omega\tau+\frac{2\pi i}{L}[|m|t(\tau)-mx(\tau)]} \frac{d\tau}{\sqrt{4\pi|m|}}.
\end{eqnarray}
Within the first-order approximation and in the interaction
picture, this evolution can be expressed as \cite{L38,L39}
\begin{eqnarray}\label{Q5}
U|g\rangle|0\rangle&=&C_{0}(|g\rangle|0\rangle-i\eta_0|e\rangle|1_{m}\rangle), \nonumber
\\U|e\rangle|0\rangle&=&C_{1}(|e\rangle|0\rangle+i\eta_1|g\rangle|1_m\rangle),
\end{eqnarray}
where  $C_{0}$ and $C_{1}$ are the normalization factors.
In this paper, we assume that the accelerated trajectory of the detector is
$t(\tau)=a^{-1}\sinh(a\tau)$ and
$x(\tau)=a^{-1}[\cosh(a\tau)-1]$ with $a$ being the proper acceleration.
Denoting $\eta_0=\lambda{\sum_m I_{+.m}}$ and $\eta_1=\lambda{\sum_m I_{-.m}}$, Eq.(\ref{Q5}) can be rewritten as
\begin{eqnarray}\label{Qq5}
\nonumber U|g\rangle|0\rangle&=&\frac{1}{\sqrt{1+|\eta_{0}|^{2}}}(|g\rangle|0\rangle-i\eta_{0}|e\rangle|1_m\rangle), \nonumber
\\U|e\rangle|0\rangle&=&\frac{1}{\sqrt{1+|\eta_{1}|^{2}}}(|e\rangle|0\rangle+i\eta_{1}|g\rangle|1_m\rangle).\nonumber
\end{eqnarray}

If the detector is initially in its ground state, then the excitation probability is given by
\begin{eqnarray}\label{QQ}
P=\sum_{m\neq0}|\langle1,e|U^{(1)}|0,g\rangle|^2=\lambda^2\sum_{m\neq0}|I_{+,m}|^2.
\end{eqnarray}
From Eq.(\ref{QQ}) we can see that the transition probability is dependent on the concrete parameters, such as the length of cavity $L$, the energy gap $\Omega$, and the interaction time scale $\sigma$.
In particular, the transition probability may decrease with the growth of acceleration when the interaction timescale $\sigma$ is much smaller than the reciprocal of the
energy gap $\Omega^{-1}$. In other words, the detector is not be warmed but cooled down.
This counterintuitive effect is called anti-Unruh effect.

\section{Quantum entanglement and coherence for the Unruh-DeWitt model  \label{GSCDGE2}}
Quantum entanglement and coherence are two important quantities for describing quantum states. They are closely related to each other but also have own characteristics.
In the case of two-qubit systems, quantum entanglement can be effectively described by the concurrence \cite{L44,L45}
\begin{eqnarray}\label{Q7}
E(\rho_{AB})=\max \{0,\sqrt{\lambda_1}-\sqrt{\lambda_2}-\sqrt{\lambda_3}-\sqrt{\lambda_4} \},
\end{eqnarray}
where $\lambda_i$ are the eigenvalues of the matrix $\rho_{AB}[(\sigma_y\otimes\sigma_y)\rho_{AB}^*(\sigma_y\otimes\sigma_y)]$ in decreasing order. On the other hand, quantum entanglement can also be measured by the logarithmic negativity $N(\rho_{AB})$, which is defined as \cite{L29}
\begin{eqnarray}\label{QQ7}
N(\rho_{AB})=\log_2||\rho_{AB}^{T_A}||,
\end{eqnarray}
where $||\rho_{AB}^{T_A}||$ is the sum of the absolute values of the
eigenvalues of the partial transpose of density matrix $\rho_{AB}$ with respect to subsystem $A$.

There are several methods to describe the coherence of quantum states, in which the measure of $l_1$ norm of coherence is maybe the simple and intuitive one.
In the given reference basis, the $l_1$ norm of coherence is defined as the sum of absolute value of all the off-diagonal elements of the system density matrix  \cite{L46}
\begin{equation}\label{Q8}
    C_{l_1}(\rho_{AB})=\sum_{{i\neq j}}|\rho_{i,j}|.
\end{equation}
One can also quantify quantum coherence by the relative entropy of coherence (REC) which is given by
\begin{equation}\label{QQ8}
C_{\rm REC} \left( \rho_{AB}  \right) = S\left( {\rho
_{\rm{diag}} } \right) - S\left( \rho_{AB}  \right),
\end{equation}
where $S(\rho_{AB})$ donates the von Neumann entropy of quantum state $\rho_{AB}$, and $\rho_{\rm diag}$ denotes the state obtained from $\rho_{AB}$ by deleting all off-diagonal elements. In this paper, we employ quantum entanglement and coherence to study the characteristics of both the Unruh effect and anti-Unruh effect.

Consider a UDW detector, which is initially entangled with a spin qubit with spin up $|\uparrow\rangle$ and spin down $|\downarrow\rangle$. The detector is placed in a vacuum cavity with length $L$. The initial state of the whole system takes the form
\begin{eqnarray}\label{Q9}
|\psi\rangle_{qDC}=(\alpha|\uparrow_q\rangle|g_D\rangle+\beta|\downarrow_q\rangle|e_D\rangle)|0_C\rangle,
\end{eqnarray}
where the real coefficients $\alpha$ and $\beta$ satisfy $\alpha^2+\beta^2=1$, and $|0_{C}\rangle$ denotes the vacuum state of the cavity field.
For convenience of description, we use the subscripts $q$, $D$, and $C$ to denote the qubit, detector, and cavity field, respectively. Now we let the detector moves with a uniform acceleration $a$ in the cavity. According to Eq.(\ref{Q5}), the state of the whole system becomes
\begin{eqnarray}\label{Q10}
|\psi\rangle_{q\bar D\bar C}&=&\frac{\alpha}{\sqrt{1+|\eta_0|^2}}|\uparrow_q\rangle|g_D\rangle|0_C\rangle-
\frac{i\alpha\eta_0}{\sqrt{1+|\eta_0|^2}}|\uparrow_q\rangle|e_D\rangle|1_C\rangle \\
&+&\frac{\beta}{\sqrt{1+|\eta_1|^2}}|\downarrow_q\rangle|e_D\rangle|0_C\rangle+
\frac{i\beta\eta_1}{\sqrt{1+|\eta_1|^2}}|\downarrow_q\rangle|g_D\rangle|1_C\rangle. \nonumber
\end{eqnarray}
Here the symbol ``bar" above D and C denotes that the states for the detector and cavity field are observed in the noninertial frame determined by the accelerated detector. Eq.(\ref{Q10}) implies that the vacuum state in the inertial frame becomes anti-vacuum observed in the noninertial frame. In other words, the UDW detector would detect the production of particles in the vacuum cavity. In the following, we will study the change of quantum entanglement and coherence induced by the acceleration effect.

Let us first calculate quantum entanglement and coherence between qubit and detector. Tracing over the cavity field modes in Eq.(\ref{Q10}), we obtain the density operator between qubit and detector as
\begin{eqnarray}\label{Q11}
\rho_{q\bar D}&=&\frac{\alpha^2}{1+|\eta_0|^2}|\uparrow_qg_D\rangle\langle\uparrow_qg_D|+
\frac{\alpha^2|\eta_0|^2}{1+|\eta_0|^2}|\uparrow_qe_D\rangle\langle\uparrow_qe_D| \\ \nonumber
&+&\frac{\beta^2}{1+|\eta_1|^2}|\downarrow_qe_D\rangle\langle\downarrow_qe_D|+
\frac{\beta^2|\eta_1|^2}{1+|\eta_1|^2}|\downarrow_qg_D\rangle\langle\downarrow_qg_D| \\ \nonumber
&+&\frac{\alpha\beta}{\sqrt{(1+|\eta_0|^2)(1+|\eta_1|^2)}}
(|\uparrow_qg_D\rangle\langle\downarrow_qe_D|+|\downarrow_qe_D\rangle\langle\uparrow_qg_D|) \\ \nonumber
&-&\frac{\alpha\beta}{\sqrt{(1+|\eta_0|^2)(1+|\eta_1|^2)}}
(\eta_0\eta_1^*|\uparrow_qe_D\rangle\langle\downarrow_qg_D|+
\eta_0^*\eta_1|\downarrow_qg_D\rangle\langle\uparrow_qe_D|).
\end{eqnarray}
Now the system consists of two objects, the inertial qubit and the accelerated detector. Employing Eq.(\ref{Q7}), we obtain the concurrence $E(\rho_{q\bar D})$ between qubit and detector as
\begin{eqnarray}\label{Q12}
E(\rho_{q\bar D})=\max \bigg\{0, \frac{2\alpha\beta(1-|\eta_0||\eta_1|)}{\sqrt{(1+|\eta_0|^2)(1+|\eta_1|^2)}},
\frac{2\alpha\beta(|\eta_0||\eta_1|-1)}{\sqrt{(1+|\eta_0|^2)(1+|\eta_1|^2)}}\bigg\}.
\end{eqnarray}
We can also use Eq.(\ref{QQ7}) to get the logarithmic negativity $N(\rho_{q\bar D})$
\begin{eqnarray}\label{QQ12}
N(\rho_{q\bar D})&=&\log_2\bigg[\sqrt{\bigg(\frac{\alpha^2|\eta_0|^2}{1+|\eta_0|^2}-\frac{\beta^2|\eta_1|^2}{1+|\eta_1|^2}\bigg)^2+
\frac{4\alpha^2\beta^2}{(1+|\eta_0|^2)(1+|\eta_1|^2)}}\\ \nonumber
&+&\frac{\alpha^2}{1+|\eta_0|^2}+\frac{\beta^2}{1+|\eta_1|^2}\bigg].
\end{eqnarray}
It is shown that the concurrence $E(\rho_{q\bar D})$ and the logarithmic negativity $N(\rho_{q\bar D})$
depend not only on the initial parameters $\alpha$ and $\beta$, but also on the
acceleration $a$, the length of cavity $L$, the energy gap $\Omega$ and the interaction time scale $\sigma$, i.e., both the acceleration effect and the setup's parameters can affect quantum entanglement.

\begin{figure}[htbp]
\centering
\includegraphics[height=1.8in,width=2.0in]{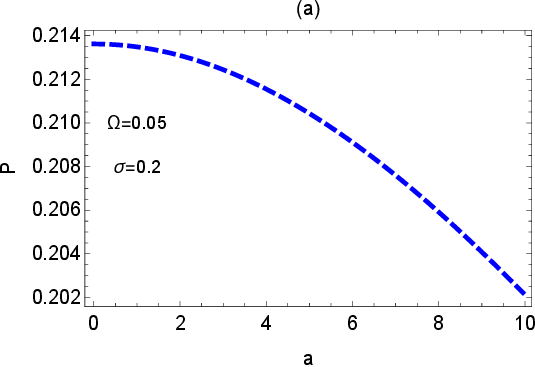}
\includegraphics[height=1.8in,width=2.0in]{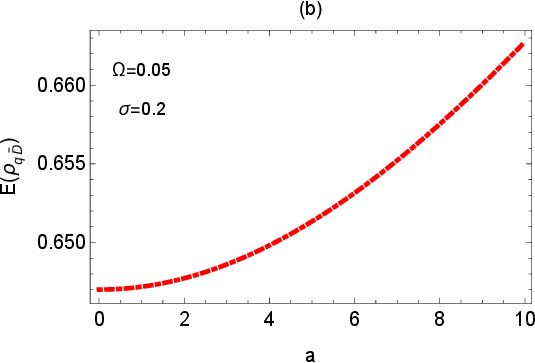}
\includegraphics[height=1.8in,width=2.0in]{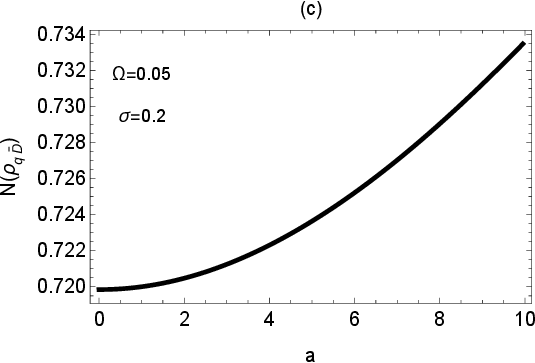}

\includegraphics[height=1.8in,width=2.0in]{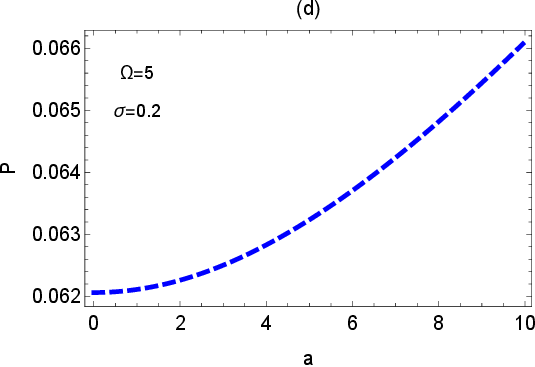}
\includegraphics[height=1.8in,width=2.0in]{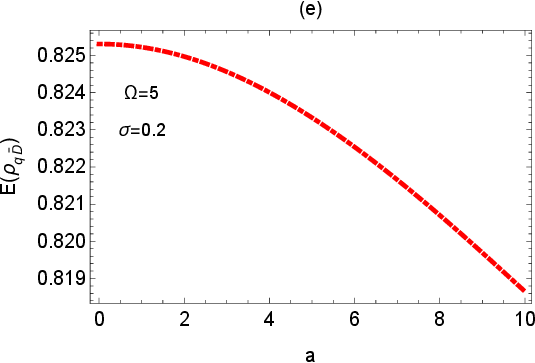}
\includegraphics[height=1.8in,width=2.0in]{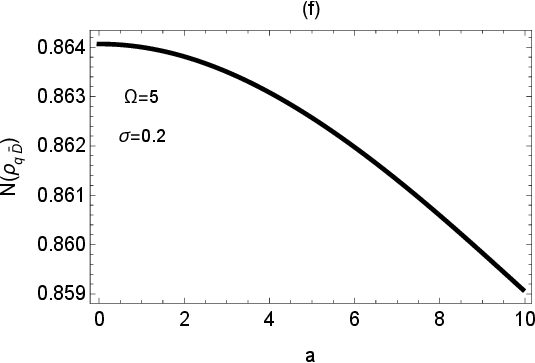}
\caption{ The transition probability $P$ (left column),  concurrence $E(\rho_{q\bar D})$ (middle column), and logarithmic negativity $N(\rho_{q\bar D})$ (right column)
 as functions of acceleration $a$ for different energy gaps $\Omega$. The rest parameters are chosen as $\alpha=\frac{1}{\sqrt{2}}$, $L=200$ and $\sigma=0.2$.}\label{Fig1}
\end{figure}

In Fig.\ref{Fig1},  we plot the transition probability $P$ (left column),  concurrence $E(\rho_{q\bar D})$ (middle column), and logarithmic negativity $N(\rho_{q\bar D})$ (right column) as functions of acceleration $a$ for different energy gaps $\Omega$, with other parameters chosen as $\alpha=\frac{1}{\sqrt{2}}$, $L=200$, and $\sigma=0.2$. It is shown that for the smaller energy gap, i.e., $\Omega=0.05$, quantum entanglement increases and the transition probability of detector decreases, with the growth of acceleration $a$, meaning that the anti-Unruh effect enhances quantum entanglement between qubit and detector. On the other hand, for the larger energy gap, i.e., $\Omega=5$, quantum entanglement decreases and the transition probability of detector increases with the growth of acceleration, showing that the Unruh effect reduces quantum entanglement between qubit and detector. Note that the initial entanglement at zero acceleration ($a=0$) is not equal to one, which is caused by the interaction between the detector and the limited vacuum field. The limited cavity space and finite interaction time $\sigma$ between the cavity field and detector lead to $\eta_{0}\neq 0$ and $\eta_{1}\neq 0$, so that the initial entanglement between qubit and detector degrades. As soon as the detector enters the cavity, it is coupled with the vacuum field in cavity, and the entanglement degradation takes place.

Besides quantum entanglement, we also study the change of quantum coherence between qubit and detector. According to Eq.(\ref{Q8}),  the $l_1$ norm of  coherence $C(\rho_{q\bar D})$ for the system of qubit and detector reads
\begin{eqnarray}\label{Q13}
C_{l_1}(\rho_{q\bar D})=\frac{2\alpha\beta(1+|\eta_0||\eta_1|)}{\sqrt{(1+|\eta_0|^2)(1+|\eta_1|^2)}}.
\end{eqnarray}
Next, we study the acceleration effect on the REC $C_{REC}(\rho_{q\bar D})$. For this purpose, we should calculate the eigenvalues of density matrix  of  Eq.(\ref{Q11}). The density matrix has two non-zero eigenvalues
\begin{eqnarray}\label{w8}
\nonumber
\lambda_1(\rho_{q\bar D})&=&\frac{\alpha^2}{1+|\eta_0|^2}+\frac{\beta^2}{1+|\eta_1|^2},\\ \nonumber
\lambda_2(\rho_{q\bar D})&=&\frac{\alpha^2|\eta_0|^2}{1+|\eta_0|^2}+\frac{\beta^2|\eta_1|^2}{1+|\eta_1|^2} \nonumber.
\end{eqnarray}
Thus, the REC of state $\rho_{q\bar D}$ becomes
\begin{eqnarray}\label{R1}
C_{REC}(\rho_{q\bar D})=\sum_{i=1}^{2} \lambda_i(\rho_{q\bar D})\log_2\lambda_i(\rho_{q\bar D})
-\sum_j \beta_j(\rho_{q\bar D})\log_2\beta_j(\rho_{q\bar D}),
\end{eqnarray}
where $\beta_j(\rho_{q\bar D})$ are the diagonal elements of $\rho_{q\bar D}$ of  Eq.(\ref{Q11}). Note that Eq.(\ref{Q11}) is a X state which has no monomeric coherence whether for the qubit or for the detector. Thus the coherence $C_{l_1}(\rho_{q\bar D})$ and $C_{REC}(\rho_{q\bar D})$ is actually a kind of genuine bipartite coherence between the qubit and the detector. According to the viewpoint of recent research \cite{AA}, this genuine bipartite coherence is actually a kind of quantum correlation between the relevant subsystems.

\begin{figure}
\begin{minipage}[t]{0.5\linewidth}
\centering
\includegraphics[width=3.0in,height=5.2cm]{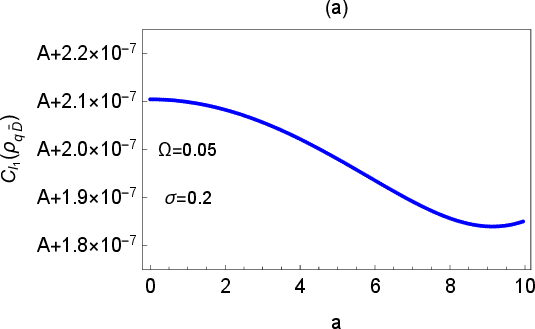}
\label{fig1a}
\end{minipage}%
\begin{minipage}[t]{0.5\linewidth}
\centering
\includegraphics[width=3.0in,height=5.2cm]{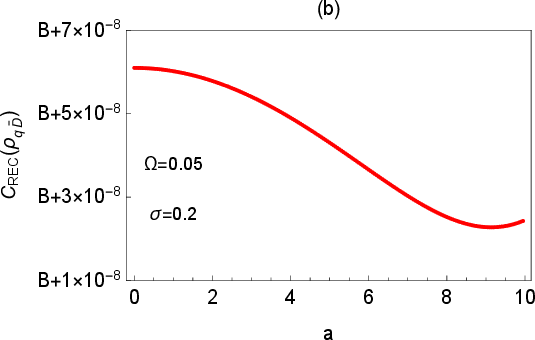}
\label{fig1c}
\end{minipage}%

\begin{minipage}[t]{0.5\linewidth}
\centering
\includegraphics[width=3.0in,height=5.2cm]{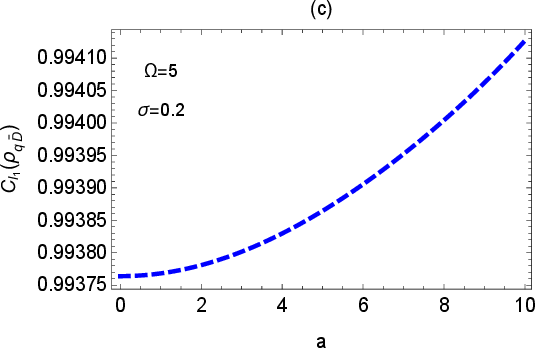}
\label{fig1a}
\end{minipage}%
\begin{minipage}[t]{0.5\linewidth}
\centering
\includegraphics[width=3.0in,height=5.2cm]{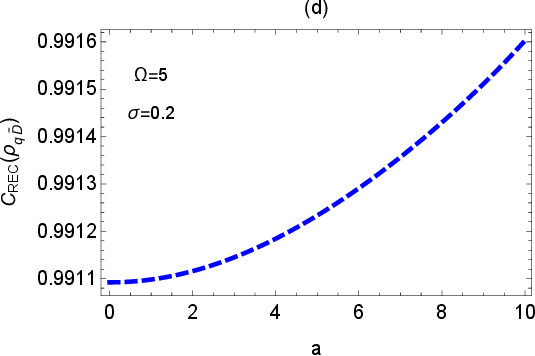}
\label{fig1c}
\end{minipage}%
\caption{The $l_1$ norm of  coherence $C(\rho_{q\bar D})$ (left column) and  the REC $C_{REC}(\rho_{q\bar D})$ (right column)
 as functions of acceleration $a$ for different energy gaps $\Omega$. The parameters are chosen as $\alpha=\frac{1}{\sqrt{2}}$, $L=200$, $\sigma=0.2$, $A=0.999999$, and $B=0.9999988$.}
\label{Fig2}
\end{figure}

For diffrent energy gaps, we plot the $l_1$ norm of  coherence $C(\rho_{q\bar D})$ and  the REC $C_{REC}(\rho_{q\bar D})$ as functions of acceleration $a$ as in Fig.\ref{Fig2}, where the parameters are chosen as the same as in Fig.\ref{Fig1}. For smaller energy gaps, i.e., $\Omega=0.05$, quantum coherence $C(\rho_{q\bar D})$ changes very slowly with acceleration. It is shown that quantum coherence $C(\rho_{q\bar D})$ decreases with acceleration $a$, meaning that the anti-Unruh effect reduces quantum coherence between qubit and detector. For bigger energy gaps, i.e., $\Omega=5$, we see that quantum coherence $C(\rho_{q\bar D})$ increases with acceleration $a$, meaning that the Unruh effect enhances quantum coherence between qubit and detector.

From above discussions, we see that the Unruh effect and anti-Unruh effect play completely opposite roles: the Unruh effect reduces quantum entanglement between qubit and detector but enhances their quantum coherence; while the anti-Unruh effect enhances quantum entanglement between qubit and detector but reduces their quantum coherence. This discovery represents a novel outcome. It was shown in the previous research that both quantum entanglement and coherence reduce under the influence of the Unruh effect \cite{L29,L30,L31,L32,L33,L34,L35,L36,SMW1,SMW2,SMW3,SMW4,SMW5,SMW6,SMW7}, which is obviously different from our results. The reason for the difference is:
the previous papers mainly consider free models in the full space and a Unruh-Dewitt model without boundary conditions, while we consider a Unruh-Dewitt model in the cavity model with boundary condition. When $a=0$, the entanglement and coherence are generally less than initial value, owing to the limited cavity space and finite interaction time  between the cavity field and detector \cite{L37,L38,L39,qsc1,qsc2}.  Physically, we can understand this phenomenon as a transfer of entanglement and coherence: as the detector enters the cavity, it interacts with the cavity mode, partially transferring entanglement and coherence from the detector to the cavity mode, thereby reducing the entanglement and coherence between detectors. An increase in acceleration may extract or transfer entanglement and coherence between the detector and cavity mode to the detectors themselves.

\section{ Conclutions  \label{GSCDGE3}}
In conclusion, we have studied the influence of acceleration effect on quantum entanglement and coherence between a qubit and a relativistic detector based on the Unruh-Dewitt model.
Depending on the chosen parameters, one can observe phenomena associated with both the Unruh and anti-Unruh effects. The Unruh effect reduces the entanglement between qubit and detector but increases their quantum coherence, challenging the notion that the Unruh effect is uniformly detrimental to quantum resources.
Contrarily, the anti-Unruh effect increases their entanglement and reduces the coherence,
indicating that the anti-Unruh effect may not always be advantageous for quantum resources.
These opposite changes between quantum entanglement and coherence under the same processes suggest the difference between them.  Furthermore, previous studies have indicated that the Unruh effect consistently leads to a simultaneous reduction in both entanglement and coherence \cite{L29,L30,L31,L32,L33,L34,L35,L36,SMW1,SMW2,SMW3,SMW4,SMW5,SMW6,SMW7}. The reason for the different results is:
the previous articles mainly consider free models in the full space and a Unruh-Dewitt model without boundary conditions, while we consider a Unruh-Dewitt model in the cavity model with boundary condition.  These results overturn conventional perceptions of the Unruh and anti-Unruh effects, simultaneously providing an unexpected source of experimental evidence for them.

\begin{acknowledgments}
The authors would like to thank Wentao Liu for helpful discussions. This work is supported by the National Natural
Science Foundation of China (Grant Nos. 12205133), LJKQZ20222315 and JYTMS20231051.
\end{acknowledgments}


\end{document}